\documentclass[english,runningheads,a4paper]{llncs}[2018/03/10]

\usepackage{regexpatch}

\makeatletter
\edef\switcht@albion{%
  \relax\unexpanded\expandafter{\switcht@albion}%
}
\xpatchcmd*{\switcht@albion}{ \def}{\def}{}{}
\xpatchcmd{\switcht@albion}{\relax}{}{}{}
\edef\switcht@deutsch{%
  \relax\unexpanded\expandafter{\switcht@deutsch}%
}
\xpatchcmd*{\switcht@deutsch}{ \def}{\def}{}{}
\xpatchcmd{\switcht@deutsch}{\relax}{}{}{}
\edef\switcht@francais{%
  \relax\unexpanded\expandafter{\switcht@francais}%
}
\xpatchcmd*{\switcht@francais}{ \def}{\def}{}{}
\xpatchcmd{\switcht@francais}{\relax}{}{}{}
\makeatother

\usepackage{newunicodechar}
\newunicodechar{°}{\degree}
\iftrue 
    \usepackage[%
      rm={oldstyle=false,proportional=true},%
      sf={oldstyle=false,proportional=true},%
      tt={oldstyle=false,proportional=true,variable=false},%
      qt=false%
    ]{cfr-lm}
\else
  \usepackage{newtxtext}
  \usepackage{newtxmath}
  \usepackage[zerostyle=b,scaled=.9]{newtxtt}
\fi

\usepackage[T1]{fontenc}
\usepackage[utf8]{inputenc} 

\usepackage{graphicx}

\usepackage{upquote}

\usepackage{booktabs}

\usepackage{paralist}

\usepackage{csquotes}

\usepackage{textcmds}

\RequirePackage[%
  final,%
  expansion=alltext,%
  protrusion=alltext-nott]{microtype}%
\DisableLigatures{encoding = T1, family = tt* }

\usepackage{url}
\makeatletter
\g@addto@macro{\UrlBreaks}{\UrlOrds}
\makeatother

\usepackage{xcolor}

\usepackage{listings}
\renewcommand{\lstlistingname}{List.}
\usepackage{chngcntr}
\AtBeginDocument{\counterwithout{lstlisting}{section}}

\usepackage{pdfcomment}
\makeatletter
\newcommand{\printfnsymbol}[1]{%
  \textsuperscript{\@fnsymbol{#1}}%
}
\makeatother

\usepackage[%
  square,        
  comma,         
  numbers,       
  sort&compress, 
]{natbib}

\SetExpansion
[ context = sloppy,
stretch = 30,
shrink = 60,
step = 5 ]
{ encoding = {OT1,T1,TS1} }
{ }

\usepackage{stfloats}
\fnbelowfloat

\usepackage{hyperref}
\hypersetup{hidelinks,
  colorlinks=true,
  allcolors=black,
  pdfstartview=Fit,
  breaklinks=true}
\usepackage[all]{hypcap}

\usepackage[group-four-digits,per-mode=fraction]{siunitx}

\usepackage{array}
\usepackage{arydshln}
\usepackage{tabularx}
\usepackage{contour}
\usepackage{ulem}

\contourlength{0.8pt}

\renewcommand{\underline}[1]{%
	\uline{\phantom{#1}}%
	\llap{\contour{white}{#1}}%
}
\newcommand{\etal}{et al.\ }

\usepackage[capitalise,nameinlink]{cleveref}
\usepackage{iflang}
\IfLanguageName{ngerman}{
  \crefname{table}{Tab.}{Tab.}
  \Crefname{table}{Tabelle}{Tabellen}
  \crefname{figure}{\figurename}{\figurename}
  \Crefname{figure}{Abbildungen}{Abbildungen}
  \crefname{equation}{Gleichung}{Gleichungen}
  \Crefname{equation}{Gleichung}{Gleichungen}
  \crefname{listing}{\lstlistingname}{\lstlistingname}
  \Crefname{listing}{Listing}{Listings}
  \crefname{section}{Abschnitt}{Abschnitte}
  \Crefname{section}{Abschnitt}{Abschnitte}
  \crefname{paragraph}{Abschnitt}{Abschnitte}
  \Crefname{paragraph}{Abschnitt}{Abschnitte}
  \crefname{subparagraph}{Abschnitt}{Abschnitte}
  \Crefname{subparagraph}{Abschnitt}{Abschnitte}
}{
  \crefname{section}{Sect.}{Sect.}
  \Crefname{section}{Section}{Sections}
  \Crefname{figure}{Fig.}{Figures}
  \crefname{figure}{figure}{figures}
  \crefname{listing}{\lstlistingname}{\lstlistingname}
  \Crefname{listing}{Listing}{Listings}
}

\usepackage{xspace}
\newcommand{\eg}{e.\,g.\ }

\renewcommand{\lastandname}{\unskip,}
\DeclareFontFamily{U}{MnSymbolC}{}
\DeclareSymbolFont{MnSyC}{U}{MnSymbolC}{m}{n}
\DeclareFontShape{U}{MnSymbolC}{m}{n}{
  <-6>    MnSymbolC5
  <6-7>   MnSymbolC6
  <7-8>   MnSymbolC7
  <8-9>   MnSymbolC8
  <9-10>  MnSymbolC9
  <10-12> MnSymbolC10
  <12->   MnSymbolC12%
}{}
\DeclareMathSymbol{\powerset}{\mathord}{MnSyC}{180}

\input glyphtounicode
\begin{document}

\title{Learning Anatomical Segmentations \index{Segmentation} for Tractography \index{Probabilistic Tractography} from Diffusion MRI}

\author{Christian Ewert\inst{1,}\thanks{These authors contributed equally.} \and
David Kügler\inst{1,}\printfnsymbol{1} \and
Anastasia Yendiki\inst{2,3} \and Martin Reuter\inst{1,2,3,}\thanks{Correspondence to \email{martin.reuter@dzne.de}}}
\authorrunning{C. Ewert et al.}

\institute{German Center for Neurodegenerative Diseases (DZNE), Bonn, Germany \and
A. A. Martinos Center for Biomedical Imaging, Massachusetts General Hospital, Boston, MA, USA \and
Department of Radiology, Harvard Medical School, Boston, MA, USA}

\maketitle              

\begin{abstract}
Deep learning \index{Deep Learning} approaches for diffusion MRI have so far focused primarily on voxel-based segmentation of lesions or white-matter fiber tracts. A drawback of representing tracts as volumetric labels, rather than sets of streamlines, is that it precludes point-wise analyses of microstructural or geometric features along a tract. Traditional tractography pipelines, which do allow such analyses, can benefit from detailed whole-brain segmentations to guide tract reconstruction.
Here, we introduce fast, deep learning-based segmentation of 170 anatomical regions directly on diffusion-weighted MR images, removing the dependency of conventional segmentation methods on T1-weighted images and slow pre-processing pipelines. 
Working natively in diffusion space avoids non-linear distortions and registration errors across modalities, as well as interpolation artifacts. \index{Registration} We demonstrate consistent segmentation results between 0.70 and 0.87 Dice depending on the tissue type. We investigate various combinations of diffusion-derived inputs and show generalization across different numbers of gradient directions. 
Finally, integrating our approach to provide anatomical priors for tractography pipelines, such as TRACULA, removes hours of pre-processing time and permits processing even in the absence of high-quality T1-weighted scans, without degrading the quality of the resulting tract estimates. 
\keywords{Diffusion MRI \and Segmentation \and Tractography \and Deep Learning}
\end{abstract}

\section{Introduction}
Tractography has significantly advanced clinical applications \cite{Costabile.2019,Shapey.2019,Essayed.2017} and has enabled neuroscientists to study developmental and pathological effects on the human connectome~\cite{Jeurissen.2019}. Traditional tractography pipelines often use anatomical segmentations to obtain priors for reconstructing tracts from diffusion-weighted MRI (dMRI). This introduces a dependency on T1-weighted (T1w) images, which are required for anatomical segmentation by neuroimaging suites such as FreeSurfer~\cite{Fischl.2012}.
For dMRI microstructural analyses, accurate segmentations of the grey/white matter (GM/WM) boundary are particularly important as different biophysical models have been proposed for each tissue type~\cite{Palombo.2020}. However, segmenting in T1w, rather than diffusion image space is problematic due to non-linear distortions between modalities, as well as potential registration inaccuracies and interpolation artifacts when mapping segmentation labels from anatomical to diffusion image space. 
Furthermore, enabled by acquisitions with high angular resolution and multiple $b$-values, dMRI-derived cytoarchitectonic boundaries may in the future complement or supersede T1w-derived segmentations for morphometric analyses.
Addressing this need for fast, accurate, dMRI-based segmentation, we present a framework for segmenting 170 GM, WM, and subcortical regions in native diffusion space, without requiring high-quality T1w images.

Methods such as SLANT and FastSurfer~\cite{Henschel.2020,Huo.2019} introduce deep learning for neuromorphometry, yet still rely on T1w images.
Traditional~\cite{Wang.2019,Yap.2015,Wang.2020,Iglesias.2019} and deep learning-based~\cite{Mu.2019,Kim.2020} methods extend segmentation to diffusion-weighted images (DWIs) for various acquisition protocols and dMRI representations. Applications of segmentation based on the inherently multi-channel dMRI signals include whole-brain GM/WM/Cerebrospinal fluid~\cite{Yap.2015,Wang.2020}, WM regions \cite{Wang.2019,Mu.2019}, nuclei (cerebellar \cite{Kim.2020} and thalamic~\cite{Iglesias.2019}),
organs~\cite{Zhang.2020,Ferreira.2020,Shehata.2019,Clark.2017}, 
tumors~\cite{Trebeschi.2017} 
and stroke lesions~\cite{NazariFarsani.2020,Clerigues.2020}.
As an alternative approach to traditional tractography, neural networks can directly segment WM tracts based on dMRI~\cite{Li.2020, Li.2019, Nelkenbaum.2020, Pomiecko.2019} or diffusion orientations~\cite{Zhang.2019, Zhang.2020b, Wasserthal.2018, Wasserthal.2019}, from clinical~\cite{Nelkenbaum.2020} or high-quality~\cite{Wasserthal.2018} datasets. Unfortunately, segmenting WM tracts as volumetric labels does not provide an along-the-tract parameterization which is useful for point-wise analyses of microstructural and geometric features of the tracts. In contrast to direct, volumetric tract segmentation approaches, the present work aims to provide the information that guides traditional tractography methods. No work to date has addressed whole-brain segmentation, including subcortical structures, cortical regions, and WM regions underlying the cortex, directly from DWIs.

\begin{figure}[t!]
\centering
\includegraphics[width=\textwidth]{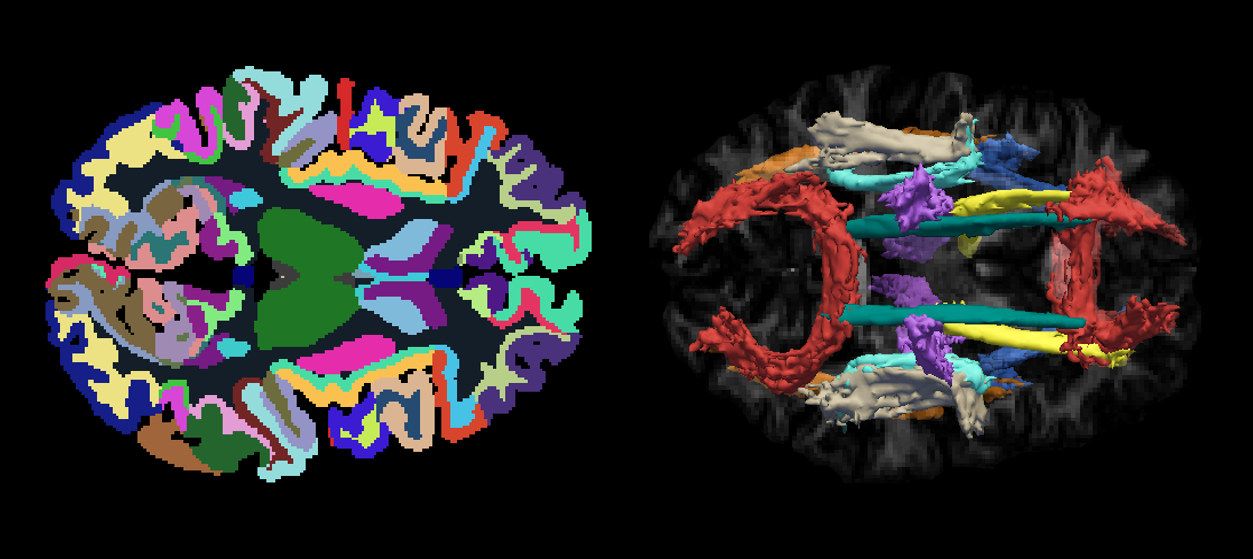}
\caption{Whole-brain segmentation of 170 regions (cortical, sub-cortical and white matter) directly from diffusion MRI (left), and probabilistic white matter tracts generated by TRACULA based on anatomical priors from dMRI-based segmentations (right)}
\label{fig:qualitative}
\end{figure}

Here we introduce deep learning-based segmentation of \underline{170 distinct regions} (cortical, subcortical, and WM) from DWIs (see Figure~\ref{fig:qualitative}). Thus we provide a fast, deep learning alternative for a critical step in dMRI pre-processing streams, which can facilitate the use of classical tractography methods, without limiting their outputs and subsequent analysis options.
Basing segmentations purely on dMRI data removes the dependency on high-quality T1w images, the potentially error-prone, cross-modal co-registration and interpolation, and avoids confounding non-linearities between anatomical and diffusion spaces.
To develop effective segmentation of dMRI data, we compile a dataset of DWIs and reference segmentations generating the latter by mapping FreeSurfer segmentations~\cite{Fischl.2012} to diffusion space.
We adopt the anatomy-targeted FastSurfer architecture~\cite{Henschel.2020}, which already supports the segmentation of a large number of regions. Moreover, expanding on the work of Li \etal\cite{Li.2020} for WM tract segmentation, we explore suitable dMRI data representations for learning-based segmentation. 
We compare inputs consisting of images without diffusion-weighting, diffusion tensor components, or DWIs and vary the number of DWIs from which tensors are generated.

When compared against FreeSurfer, our method achieves performance comparable to the state-of-the-art at orders of magnitude faster processing times.
As a use case, we integrate dMRI-based segmentations into the tractography package TRACULA (TRActs Constrained by UnderLying Anatomy)~\cite{Yendiki.2011}, which performs global probabilistic tractography with anatomical priors.  
Differences of TRACULA tracts based on deep learning vs.\ traditional anatomical initialization (see Figure~\ref{fig:tracts}) are in the order of previously reported differences between automated TRACULA reconstructions and manual labels while reducing the TRACULA pre-processing run-time by 4 hours. With our approach, the segmentation of 170 regions in dMRI space can be achieved in 32 seconds on a GPU.

\section{Materials and Methods}

The tract posterior probabilities computed by TRACULA involve anatomical priors that are calculated from an anatomical segmentation. We perform this segmentation directly in diffusion space and remove the requirement for T1w images by replacing the corresponding component of the TRACULA pipeline with a deep learning network.

\subsection{Data}
\subsubsection{Diffusion MRI data.} We use pre-processed DWIs from the WU-Minn Human Connectome Project (HCP)~\cite{vanEssen.2012, Moeller.2010, Feinberg.2011, Setsompop.2012, Sotiropoulos.2013,Xu.2012}, which are already corrected for eddy-currents and subject motion. 
These images are acquired with a 3-shell protocol at $b$-values of 1000, 2000, and 3000 \si{\second/\milli\meter^2}. 
Each shell is composed of 90 diffusion-encoding gradient directions, approximately uniformly distributed on the sphere. 
In addition, 18 images without diffusion-weighting are interleaved with the DWIs. 
For training, validation, and test, we create gender-balanced non-intersecting subsets of 250, 50, and 100 subjects, respectively. 

\subsubsection{Segmentation Labels.} Anatomical segmentations of T1-weighted images of the same subjects are obtained with FreeSurfer 6.0 \cite{Fischl.2012}. We project cortical parcellations from the surface models up to \SI{2}{\milli\meter} deep into the WM, as required by TRACULA. The registration to the diffusion space is performed with the boundary-based rigid registration method \textit{bbregister} \cite{Greve2009}. 

\subsection{Data Representations}
Since $q$-space sampling schemes may comprise anywhere from six to several hundred measurements, a general segmentation approach should be independent of the exact choice of diffusion-encoding directions and $b$-values. Instead, a suitable representation has to abstract from acquisition details yet contain sufficient relevant information. A parsimonious model that is often fitted to DWIs acquired on shells is the diffusion tensor~\cite{Basser.1994}, which models local diffusion as a single (uni-modal) Gaussian distribution. The symmetric $3 \times 3$ diffusion tensor \index{Diffusion Tensor} can be understood as a condensed \textit{summary} of the local diffusion behavior at a given voxel and can be reconstructed from any $q$-shell acquisition scheme that includes at least 6 directions. To explore how the number of DWIs used to fit the tensor affects its performance for the segmentation task, we extract multiple subsets of gradient directions on the same shell (approximately uniformly distributed). For each of these subsets, the diffusion tensor is fitted to the data with \textit{FSL}'s \textit{dtifit} function and the six unique tensor components are stacked and used as input.

\subsection{Architecture}

FastSurferCNN \cite{Henschel.2020} is an U-Net-based neuroimage segmentation network validated extensively on anatomical MRI datasets. Three fully convolutional networks \index{Convolutional Neural Network} are trained independently on axial, coronal, and sagittal slices of MR images. The predictions from the three views are then combined into the final prediction volume by means of a weighted average (view-aggregation). The network uses skip-connections between encoder- and decoder-blocks. In the decoder-blocks, information from the previous decoder-block and the corresponding encoder-block are combined. Instead of simply concatenating these feature maps, \mbox{FastSurferCNN} employs competitive dense blocks to reduce the number of parameters. Competitive dense blocks rely on max-out activations to encourage the network to learn which parts of the provided feature maps are relevant for the segmentation task.

All networks are trained with a combined loss-function containing a median frequency balanced logistic loss with edge-focus and a Dice loss:

$$\mathcal{L} = \underbrace{-\sum_x \omega(x)g_c(x)log(p_c(x))}_{\text{Logistic Loss}}-\underbrace{\frac{2\sum_x p_c(x) g_c(x)}{\sum_x p_c^2(x)+\sum_x g_c^2(x)}}_{\text{Dice Loss}}$$

with $\omega(x) = \omega_F(x)+\omega_E(x)$, median frequency balanced weights $\omega_F(x)$, edge-weighting $\omega_E(x)$ at voxel $x$, references $g$, prediction $p$ and class $c$. In order to provide 3D spatial context, FastSurferCNN's input consists not only of the slice of interest but also a sequence of neighboring slices.

\subsection{Training}
We train all networks with an initial learning rate of $0.01$, which is reduced every 10 epochs (multiplied by $0.2$). Early stopping is applied when the loss on the validation set does not improve for 15 epochs.

\subsection{Tracts}
To calculate priors for tractography, TRACULA needs registrations between anatomical, diffusion and MNI spaces. Since we provide the segmentation natively in diffusion space, a registration to anatomical T1w space is obsolete and a single registration from diffusion to MNI space suffices. This mapping can easily be established without a T1w image, yet the two different registration-methods (one includes the anatomical image whereas the other does not) introduce an additional variance which prevents a consistent evaluation of tract similarity. We, therefore, re-use the mapping from diffusion to MNI space that was established to generate the reference tracts. Furthermore, to mitigate artifacts from nearest-neighbor interpolation to large voxel sizes, we instead predict and map probability distributions over classes (soft-labels) via tri-linear interpolation, taking the argmax over classes in the target space.

\subsection{Evaluation Criteria}
\subsubsection{Segmentation Quality.}
We measure the similarity of the segmentation resulting from our method and the segmentation of FreeSurfer mapped to diffusion space, with two evaluation criteria.

The \emph{Dice score $\mathcal{D}_c$} for region $c$ measures the relative overlap between the binary labels of the prediction $P_c = \{p_{ic} \mid i=1,...,N\}$ and the reference $R_c=\{r_{ic} \mid i=1,...,N\}$ segmentation:

$$\mathcal{D}_c(P, R) = \frac{2 \sum_{i=1}^N p_{ic} r_{ic}}{\sum_{i=1}^N p_{ic} + \sum_{i=1}^N r_{ic}}$$.

The voxel-based mean \emph{Hausdorff distance $\mathcal{H}_c$} for region $c$ measures the difference between prediction $P_c$ and reference labels $R_c$ via 

$$\mathcal{H}_c(P, R) = \frac{1}{|R_c|}\sum_{r\in R_c} \min_{p \in P_c} ||p-r||_2 + \frac{1}{|P_c|}\sum_{p \in P_c} \min_{r \in R_c} ||p-r||_2$$.

\subsubsection{Tractography.}

Similarly to the segmentation case, we quantify the similarity between pairs of tracts reconstructed with anatomical priors from either segmentation via voxel-based mean \emph{Hausdorff distance}. Since TRACULA relies on a Markov-Chain Monte-Carlo method, two different sets of tracts are not \textit{per se} comparable. Thus, in accordance with previous tract evaluations~\cite{Yendiki.2011, Zollei.2019}, we threshold the tracts at 20\% of their maximum intensity.

\section{Results and Discussion}

To determine the impact on segmentation quality, we evaluate networks with respect to different data subsets and input representations, keeping the network architecture (\eg number of filters) fixed. 

\subsection{Evaluation 1: $Q$-Space Sampling Density}
\label{sec. 2}
\begin{figure}[h!]
\centering
\fcolorbox{black}{white}{\includegraphics[width=0.975\textwidth]{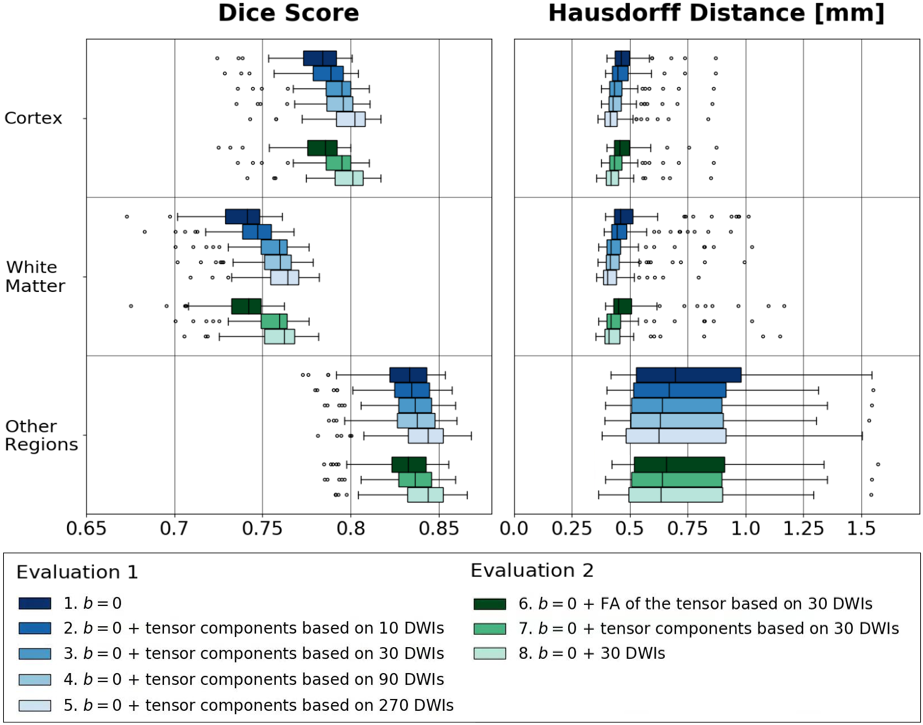}}
\caption{Ablation of neural network input in comparison to FreeSurfer reference segmentation. \textbf{Evaluation~1}: $Q$-Space Sampling Density (top group of blue bars): 1.~Only $b=0$ image, 2-5.~$b=0$ + diffusion tensors fitted with varying sampling density (2-4.~only on first shell with $b=\SI{1000}{\second/\milli\meter^2}$, 5.~on three shells); \textbf{Evaluation~2}: dMRI data representation (bottom group of green bars) of a fixed set of 30 DWIs on the first shell: $b=0$ plus 6.~an FA map, 7.~the diffusion tensor, and 8.~DWIs directly.}
\label{fig:directions}
\end{figure}

When $q$-space is sampled more densely, we expect the diffusion tensor to be more accurate due to the improved signal to noise ratio (SNR). To explore how different single-shell $q$-space samplings influence the segmentation quality, we compare FreeSurfer segmentations against our network's prediction for several scenarios which are displayed in Figure \ref{fig:directions}.

The Dice score seems to correlate with the compactness of the shape of anatomical regions. Scores are high for sub-cortical regions, where shapes feature large volume-to-surface ratios, lower in the folded cortical areas, and lowest for the thin cortical projections into the WM. The slim, \SI{2}{\milli\meter} projection on a coarse voxel grid with \SI{1.25}{\milli\meter} isotropic size could be another reason for the smaller Dice scores in WM regions. In fact, the ratio of voxels on the region boundary is significantly higher for white matter regions compared to cortical regions. The Hausdorff distances paint a very similar picture in terms of ranking methods while they are more consistent than Dice scores across WM- and cortical regions.

As expected, the segmentation quality increases when the diffusion tensor is fitted with more DWIs (1.-5.). The segmentation based solely on the image without diffusion-weighting ($b=0$) provides a strong baseline, potentially due to the higher SNR at $b=0$. Yet, the inclusion of additional diffusion information increases segmentation performance further. 

\subsection{Evaluation 2: Input Representations}

We also assess the effect of different input data on the segmentation quality (see Figure \ref{fig:directions}, second, green bar group: 6-8.).
\textit{Fractional Anisotropy (FA)} \index{Fractional Anisotropy} measures the coherence of water diffusion in a voxel and is frequently used in tract-based analyses. Since \textit{FA} is a scalar measure computed from the eigenvalues of the tensor, it only contains a subset of the tensor information. As a result, the performance is worse with \textit{FA} (6.) than with the full tensor (7.).
More broadly, the diffusion tensor is a simple model that fails to accurately describe water diffusion in full detail. Thus, it is not surprising that a segmentation directly based on the DWIs (8.) yields better results than one based on the diffusion tensor. However, for the task of segmentation, the diffusion tensor seems to capture most of the relevant information present in the DWIs.

\subsection{Evaluation 3: Generalization}

\begin{figure}[h!]
\centering
\fcolorbox{black}{white}{\includegraphics[width=0.975\textwidth]{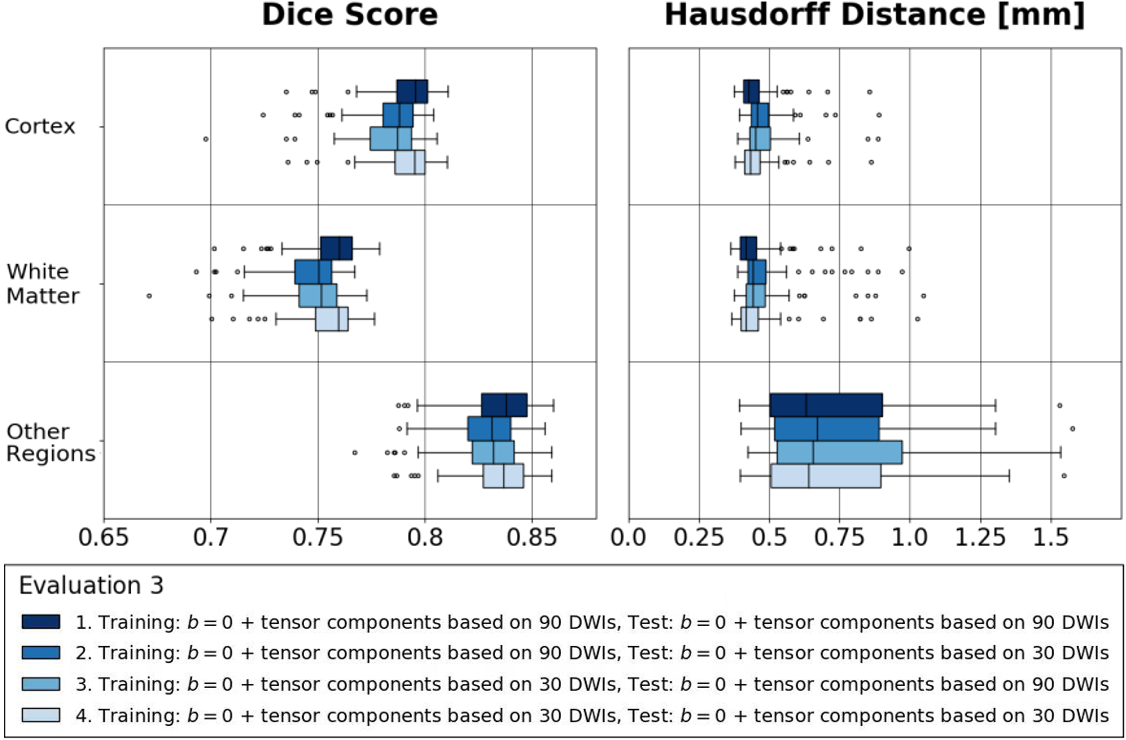}}
\caption{Network generalization when tensors are fitted with differently many DWIs.}
\label{fig:cross}
\end{figure}

In the previous evaluation, networks were trained separately for each set of inputs. This evaluation explores how a network \textit{trained} on tensor components based on $n$ DWIs performs when it is \textit{evaluated} on tensor components based on $m$ DWIs (with $n \neq m$). This kind of generalizability is a critical property when accommodating data with a variety of acquisition details at test time. Notably, while neither network generalizes perfectly to the different input format, the stable results (Figure~\ref{fig:cross}) suggest that generalizability can be asserted to a large extent.

\subsection{Evaluation 4: Tract Similarity}

\begin{figure}[h!]
\centering
\fcolorbox{black}{white}{\includegraphics[width=0.975\textwidth]{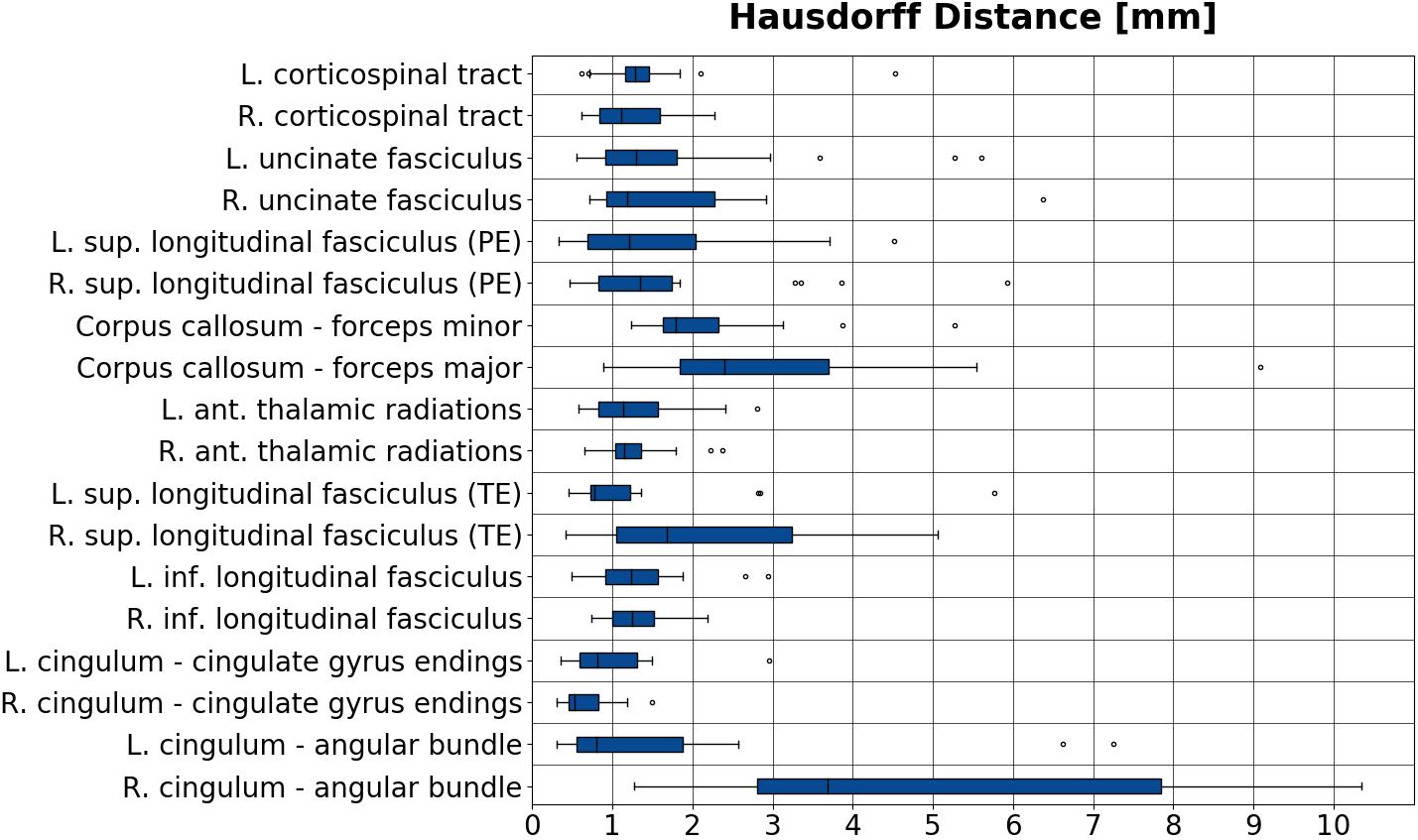}}
\caption{Similarity of tracts based on DWI segmentations versus FreeSurfer's segmentation. L.=Left, R.=Right, sup.=superior, ant.=anterior, inf.=inferior, TE=temporal endings, PE=parietal endings}
\label{fig:tracts}
\end{figure}

Finally, we assess the stability of WM tract generation when switching from traditional T1w image segmentation to our dMRI-based, deep learning approach:
For a gender-balanced subset of 20 subjects from the HCP, we reconstruct 18 different WM tracts with TRACULA, using either the proposed, dMRI-based segmentations or the FreeSurfer T1w-based segmentations (see Figure \ref{fig:tracts}). The deviation is within the margin of the deviation of TRACULA's tracts from manually-annotated bundles~\cite{Yendiki.2011}, which is around \SI{2}{\milli\meter} for most tracts. 

We time both applications on five representative cases (Evaluation 1 and 4). Our method takes \SI{32}{\second} for the anatomical segmentation in diffusion space compared to \SI{4}{\hour} with FreeSurfer (parallelization of hemispheres and 4 threads, \SI{7}{\hour} sequentially). For the tractography pipeline, our work accelerates the total run time from \SI{283}{\minute} to \SI{56}{\minute}.

\section{Conclusion}

Our work presents and analyzes the application of deep learning for anatomical segmentation directly on DWIs.
Applied to probabilistic tractography with anatomical priors, our method enables processing without the requirement of T1w images and thus avoids errors from non-linear distortions, registration inaccuracies, or interpolation artifacts.
As a consequence, dMRI-based anatomical segmentation achieves results similar to corresponding state-of-the-art T1w-based segmentation and speeds up pre-processing in the TRACULA pipeline by hours. Accelerating heavy processing pipelines is essential especially for large cohort studies such as HCP~\cite{vanEssen.2012} or the Rhineland study~\cite{Breteler.2014}, where large diffusion datasets from thousands of participants require efficient methods. 

Furthermore, we analyze how the choice of input images for the neural network affects segmentation performance.
We confirm that segmentation quality increases as more $q$-space samples are included when fitting diffusion tensors -- likely due to increased SNR.
Skipping the tensor fit and directly learning from DWIs increases segmentation performance further.
However, simply increasing $q$-space samples is not an option due to memory limitations and reliance on the availability of the same set of $q$-space samples for future input cases. 
Tensor-based inputs, on the other hand, provide a widely applicable alternative~\cite{Li.2020} and yield results that remain relatively stable and close to the direct DWI performance. 
In our opinion, tensors fitted to 30 DWIs, a number that is feasible in clinical studies, offer a good balance. 
Future work will explore diffusion models other than the tensor as segmentation inputs and assess generalizability across a large variety of different dMRI datasets.

While we illustrate the use of our deep learning-based segmentation in a pipeline for probabilistic tractography with anatomical priors, it can be useful in a wide range of other applications. These include improved WM/GM and pallidum-putamen segmentation, seed-based tractography, network analysis, or ROI-based analysis of microstructural measures, to name a few.

\section{Acknowledgements}

Data were provided in part by the Human Connectome Project, WU-Minn Consortium (Principal Investigators: David Van Essen and Kamil Ugurbil; 1U54MH091657) funded by the 16 NIH Institutes and Centers that support the NIH Blueprint for Neuroscience Research; and by the McDonnell Center for Systems Neuroscience at Washington University.
AY was partly supported by NIH awards R01-EB021265 and U01-EB026996.

\renewcommand{\bibsection}{\section*{References}} 
\bibliographystyle{splncsnat}
\begingroup
  \ifluatex
  \else
    \microtypecontext{expansion=sloppy}
  \fi
  \small 
  \bibliography{references}
\endgroup
%
\end{document}